\documentstyle[11pt,epsfig]{article}

\newcommand{\lwig}{\mbox{\,\raisebox{.3ex}
{$<$}$\!\!\!\!\!$\raisebox{-.9ex}{$\sim$}\,}}
\newcommand{\gwig}{\mbox{\,\raisebox{.3ex}
{$>$}$\!\!\!\!\!$\raisebox{-.9ex}{$\sim$}}\,}

\def\ai           {{\overline{I}}}
\def\i            {I}
\def\iai          {\mbox{$I\overline{I}$}}
\def\ii            {{\rm i}}

\font\tenrm=cmr10

\newcommand{\capfont}{\tenrm}

\begin{document}
\date{}
\title{\normalsize{\rightline{DESY 94-197}\rightline{hep-ph/9411217}}
    \vskip 1cm
     \bf Towards the Phenomenology of  \\
               \vglue 0.2cm
    QCD--Instanton Induced Particle Production \\
               \vglue 0.2cm
    at HERA\thanks{to pe published in Proc. of the International Seminar
``Quarks--94'', Vladimir/Russia, May 11-18, 1994}
\\
\vspace{11mm}}
\author{ A. Ringwald and F. Schrempp\\[5mm]
{ Deutsches Elektronen-Synchrotron DESY, Hamburg, Germany}\\ \vspace{2mm}}
\begin{titlepage}
\maketitle
\begin{abstract}
We present a first status report on a
broad and systematic study of possible manifestations
of QCD-instantons at HERA.
Considerable motivation comes from the close analogy between
instanton-induced $B+L$ violation in electroweak processes and
effects of QCD-instantons in deep inelastic scattering.
We concentrate on the high multiplicity final state structure,
reminiscent of an isotropically decaying ``fireball''.
A set of experimental
isolation criteria is proposed. They serve to further
enhance the striking event signature without significantly
suppressing the expected rates.
\end{abstract}

\thispagestyle{empty}
\end{titlepage}
\newpage
\setcounter{page}{2}
\section{Introduction}
The basic significance and possible importance of QCD-instanton
effects in deep inelastic scattering for decreasing  Bjorken variable
$x_{\rm Bj}$ and high photon virtuality $Q^2$
has recently been emphasized \cite{bb}.

First of all, it  has been argued \cite{bbgg,bb} that the calculable
small scale instanton dynamics \cite{kr} may be factorized
from the large distance effects, allowing for a semiquantitative
estimate of instanton-induced contributions
to the structure functions \cite{bb} and the hadronic final state \cite{rs}.
This is rooted in the fact that $1/\sqrt{Q^2}$ plays the r\^ole of
a dynamical infrared cut-off for the instanton size \cite{kr}.

Secondly, QCD-instanton effects for decreasing $x_{\rm Bj}$ are
largely analogous to the manifestation
of electroweak  instantons for increasing energies \cite{kr}.
The anomalous $B+L$ violation due to electroweak instantons is paralleled by a
chirality violation induced by QCD-instantons \cite{th}.
The spectacular possibility of a strong increase of
multi-$W^\pm /Z^0$ production
in the multi-TeV regime \cite{r,m} due to electroweak instantons
corresponds to a strong enhancement of  multi-gluon  production
at small $x_{\rm Bj}$ due to QCD-instantons. Striking consequences
include a high multiplicity final state structure,
reminiscent of a decaying ``fireball''.

Whereas a promising search for anomalous
electroweak events is only possible in the far future, presumably
at a post-LHC collider \cite{ewcollpheno} or at
cosmic ray facilities
\cite{ewcosmray}, the search for anomalous events induced
by QCD-instantons can start right now,
in deep inelastic $e^\pm p$ scattering at HERA.

The present paper represents a first status report on a
broad and systematic study of possible manifestations
of QCD-instantons which could be searched for at HERA \cite{rs}.
While a theoretical derivation of crucial quantities
characterizing the instanton-induced final state ($1,2\ldots$ particle
inclusive rates, average multiplicity, average transverse momenta,
energy flow, etc.)
will be deferred to Ref.~\cite{rs}, we mainly report here on
the phenomenological aspects of our results.

The organization of this contribution is as follows.
We start off in Section~2 by expanding on the close analogy between
instanton-induced $B+L$ violation in electroweak processes and
effects of QCD-instantons in deep inelastic scattering.
Section~3 contains a summary of the results of Balitsky and Braun \cite {bb}
concerning the QCD-instanton induced contribution to the
gluon and quark structure functions.
We also emphasize the  approximations and limitations inherent
in this calculation. Furthermore,
we present the instanton contribution to the nucleon structure
function $F_2(x_{\rm Bj},Q^2)$ obtained by convoluting the results
of Ref.~\cite{bb} with phenomenological distributions
of quarks and gluons in the nucleon.
In Section~4 we report on our ongoing investigation \cite{rs}
of the QCD-instanton induced hadronic final state in
deep inelastic scattering. The main emphasis rests on the
characteristic event topology along with a discussion of experimental
isolation criteria, serving to further
enhance the striking event signature without significantly
suppressing the rates. A  search strategy for
instanton-induced events is formulated. In Section~5, we present
a summary and an outlook on related aspects and open problems under study.

\vspace{0.2cm}
\section{The QFD -- QCD Connection}
The Standard Model of electroweak (QFD) and strong (QCD)
interactions is remarkably successful.
In particular, its perturbative
formulation (``Feynman diagrammatics")
appears to be theoretically consistent
and agrees with precision experiments
(where applicable, i.e. for
small coupling constant).

Nevertheless, even for small couplings, there exist physical processes
which cannot be described by conventional perturbation theory,
notably, phenomena associated with {\it quantum
tunnelling}.

\begin{figure}
\vspace{-0.4cm}
\begin{center}
\epsfig{file=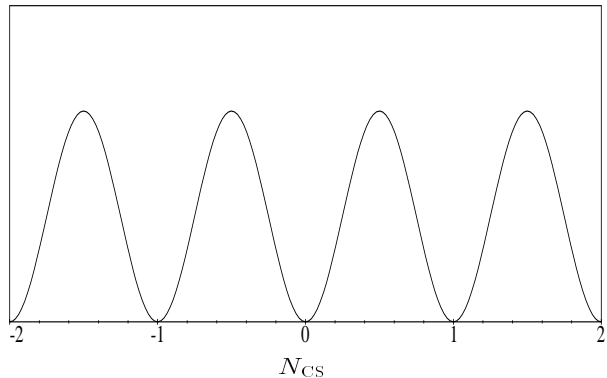,bbllx=126pt,%
bblly=280pt,bburx=489pt,bbury=564pt,width=8cm,height=5cm}
\caption[dum]{\capfont \label{f1}
Schematic illustration of the static
potential energy of the gauge (and Higgs) field vs.
the Chern-Simons number}
\end{center}
\end{figure}

In non-Abelian gauge theories such as QFD and QCD
the vacuum actually has a complicated structure, even on the
classical level \cite{jr}: The potential energy is periodic
with respect to the
so-called Chern-Simons number (c.\,f. Fig.~\ref{f1}),
\begin{equation}
N_{\rm CS} [A] = {g^2\over 16\pi^2}
\int d^3x\ \epsilon_{ijk} \biggl(
A_i^a\partial_j A_k^a - {g\over 3} \epsilon_{abc}
A_i^a A_j^b A_k^c \biggr)  ,
\end{equation}
which is
the (topological) winding number of the (non-Abelian) gauge field ($A$)
under consideration.
Pure gauge fields corresponding to
the degenerate minima of the potential
energy (perturbative vacua) have integer values of the
Chern-Simons number $N_{\rm CS}$.
Moreover, pure gauge fields differing by
$\triangle N_{\rm CS}=n={\rm integer}$
are topologically inequivalent. They are related to each other
by a topologically non-trivial static gauge transformation with
winding number $n$. This means that they
 are separated by an energy barrier, as shown
schematically in Fig.~\ref{f1}.

In the electroweak theory
a mass scale, $v\approx 246$ GeV, is introduced via
spontaneous symmetry breaking and, correspondingly,
there is a definite
minimum barrier height
associated with the $W$-mass, of order
\cite{km}
\begin{equation}
\label{ewsp}
M_{\rm barrier}^{\rm QFD} \sim {m_W\over \alpha_W} \sim {\cal O}(10\ {\rm
TeV}) .
\end{equation}
As is well known, this minimum
barrier energy is associated with a certain static, unstable solution to
the classical field equations, the so-called `sphaleron' \cite{km}.
It may be viewed \cite{ag} as an intermediate,
coherent field configuration consisting
of a large number, ${\cal O}(1/\alpha_W )$, of
$W/Z$ (and Higgs) bosons, confined to a characteristic
volume of order $m_W^{-3}$.

In QCD, being an ``unbroken'' gauge theory, the minimum barrier height
depends on the considered process and its kinematics.
It turns out, that the notion of the minimum
barrier may be qualitatively transcribed from QFD,
provided that there is a (process dependent) hard scale $Q^\prime$ available.
Specifically, in deep inelastic $e^\pm p$ scattering,
\begin{equation}
Q^{\prime\,2}\propto Q^2= -({\rm mass})^2\ \mbox{ of the virtual
photon} .
\end{equation}
Besides $\alpha_W \rightarrow \alpha_s$, one is then led to substitute
the characteristic linear dimension \cite{kr,bbgg,bb,bbjet}
\begin{equation}
\label{subst}
{1\over m_W} \to {1\over \alpha_s(Q^\prime ) Q^\prime } .
\end{equation}
The minimum barrier energy ( from Eqs.~(\ref{ewsp},\,\ref{subst}))
\begin{equation}
\label{QCDsp}
M_{\rm barrier}^{\rm QCD} \sim Q^\prime ,
\end{equation}
is now associated with a sphaleron-like intermediate, coherent field
configuration consisting of a large number, ${\cal O}(1/\alpha_s )$, of
gluons in a characteristic volume $(\alpha_s Q^\prime )^{-3}$.

Transitions between minima of the effective potential
in Fig.~\ref{f1} lead to
a violation of fermionic quantum numbers in the Standard Model \cite{th,ano}.
In particular, baryon ($B$) and lepton ($L$) number conservation
is violated
due to non-perturbative electroweak gauge fields ($W$)
associated with the SU(2) flavour gauge group, according to
the selection rule
\begin{equation}
\label{anomew}
\triangle L_{e,\,\mu ,\,\tau }={1\over 3} \triangle B =
-\triangle N_{\rm CS}[W] .
\end{equation}
In analogy, non-perturbative gluon fields ($G$)
associated with the SU(3) colour gauge group
induce a violation of  chirality conservation for (massless)
quarks,
\begin{equation}
\label{anomqcd}
\triangle Q^5_{u,\, d,\, s,\,\ldots } = 2\ \triangle N_{\rm CS}[G] .
\end{equation}

For parton-parton center of mass
(c.m.) energies\footnote{
Henceforth, we shall denote by $\sqrt{s^\prime }$
the total c.m. energy of the non-perturbative subprocess,
while $\sqrt{s}$ refers to the total c.m. energy
of the physical process under consideration.}
 $\sqrt{ s^\prime }< M_{\rm barrier}$,
such processes are classically forbidden and only occur via quantum tunnelling
under the barrier in Fig.~\ref{f1}.
In this case, the sphaleron-like intermediate
state corresponding to the energies
(\ref{ewsp}), (\ref{QCDsp}) can
only be reached virtually.
Correspondingly, the respective
cross sections are exponentially suppressed in the coupling
\cite{th},
\begin{equation}
\label{tunnelling}
\sigma_{\rm tunnelling}\sim \exp (-4\pi /\alpha ) \,;\hspace{0.5cm}
\alpha =\alpha_W,\alpha_s .
\end{equation}

Let us recall the origin of this tunnelling suppression factor
in somewhat more detail.
The amplitude of anomalous fermion number violating processes
can be obtained by expanding the  path integral for
the corresponding Green's functions about {\it instantons}
\cite{bpst,th,a},
which are
classical solutions of the Yang-Mills(-Higgs) equations in
4-dimensional Euclidean space with finite
action. The instanton ($I$) (anti-instanton ($\ai$))
may be viewed as a most probable tunnelling
solution, interpolating in Euclidean time between the gauge (and Higgs) field
configurations of two neighbouring vacua with
$\triangle N_{\rm CS} = +1(-1)$. It passes the above-mentioned
sphaleron-like intermediate state in-between.

In QCD, for example, the instanton is
explicitly given by\footnote{
We use here the notations $\sigma_\mu^{\alpha\dot\alpha} = (-i\sigma, 1)$,
$\bar\sigma_{\mu\dot\alpha\alpha} = (+i\sigma, 1)$
($\sigma$ are the standard  Pauli matrices). Furthermore we
abbreviate $x = x_\mu\sigma_\mu , \bar x = x_\mu{\bar\sigma}_\mu$, etc. }
\begin{equation}
\label{eval}
G^{(I)}_{\mu}(x-x_I;U_I,\rho_I)
 =
-{\ii\over g} {U_I[\sigma_{\mu} (\bar x-{\bar x}_I) -(x-x_I)_{\mu}]
{\bar U}_I\over (x-x_I)^2
((x-x_I)^2+\rho_I^2)}
\rho_I^2  ,
\end{equation}
depending on a set of collective coordinates
$\{ x_I,\rho_I, U_I\}$, like center, $x_I$, size, $\rho_I$,
and orientation in group space, $U_I$. Since the action is independent
(QCD) or only slightly depends (QFD) on these collective coordinates,
they are to be integrated over. For the simplest exclusive
anomalous processes one obtains (for QFD, see Refs.~\cite{th,r},
for QCD, see Refs.~\cite{kr,bbgg,bb})
in this way to exponential
accuracy:

\begin{flushleft}$\triangle (B+L)=-2 n_{\rm family}=-6:$
\end{flushleft}
\begin{equation}
T\,(qq\to 7\overline{q}3\overline{l}\,)
\propto \int d\rho_I\
\ ...\
\exp\biggl[ -{2\pi\over \alpha_W(\rho_I^{-1})} S_I^{\rm QFD}(\rho_I )\biggr] ,
\label{simpleampl}
\end{equation}
\begin{flushleft}
$\triangle Q^5=2 n_f=6:$
\end{flushleft}
\begin{equation}
\left\{
\begin{array}{lll}
T\,(g^\ast g &\rightarrow & 3q_L 3\overline{q_R})\\
T\,(\overline{q_L}^\ast g &\rightarrow & 2q_L 3\overline{q_R})\\
T\,(\overline{q_L}^\ast q_R & \rightarrow & 2q_L 2\overline{q_R})\\
                       &\cdots&
\\
\end{array}\right\}
\propto  \int d\rho_I\
\ ...\
\exp\biggl[ -Q^\prime \rho_I -
{2\pi\over \alpha_s(\rho_I^{-1})} S_I^{\rm QCD}\biggr] ,
\label{simpleamplqcd}
\end{equation}
where, in our normalization, the action of the QCD-instanton
is given by   $S_I^{\rm QCD}=1$, whereas  the action of
the QFD-instanton  reads
$S_I^{\rm QFD}=1+(1/2)\rho_I^2m_W^2$.
In the various QCD-instanton induced subprocesses of
Eq.~(\ref{simpleamplqcd}),
the star indicates that the corresponding parton
carries a virtuality $Q^\prime =\sqrt{-q^{\prime\,2}}>0$.
We observe that,
for exclusive anomalous processes, the
instanton size is effectively cut off at $\rho_{\rm cut} \sim v^{-1}$
in the electroweak theory \cite{th,a} and  at $\rho_{\rm cut} \sim Q^{
\prime\,-1}$
 in QCD \cite{kr,bbgg,bb}
(for early discussions, see Ref.~\cite{earlier}).

Most interesting  from a theoretical point of view and
in the light of present and future collider experiments is the
case $\sqrt{s^\prime } > M_{\rm barrier}$, where
a transition over the barrier
is classically allowed, i.e. energetically possible. Unfortunately,
the crucial {\it dynamical} question is still unsettled \cite{m}
whether the transition from a state with a few initial partons to the {\it very
different}  sphaleron-like multi-parton coherent state can proceed without
extra suppression. Only then  could such anomalous processes
acquire an observable cross section and the final state would
consist of  a  large number,
${\cal O}(\alpha_{W(s)}^{-1})$, of $W/Z$'s (gluons)
in addition to the few fermions required by the anomaly.

The possibility that this intriguing scenario
might be  realized in nature was first observed in the context
of the electroweak theory in Ref.~\cite{r}.
It was found \cite{holy} that, to exponential accuracy, the
total cross section for anomalous $B+L$ violation,
in the high energy
and weak coupling limit,
\begin{equation}
\frac{\sqrt{ s^\prime }}{m_W}\rightarrow \infty,\hspace{1cm}
\frac{\alpha_W\sqrt{ s^\prime }}{m_W}\ \mbox{ fixed},
\label{hewc}
\end{equation}
can be written in the following scaling form
\begin{equation}
\label{totew}
\sigma^{(I)\,{\rm tot}}_{\rm QFD} \propto
\exp\left[ -{4\pi\over \alpha_W(\rho_\ast^{-1})}\ F^{\rm QFD}\left(
{\sqrt{ s^\prime }\over M_0} \right) \right] ,
\end{equation}
where $M_0=\sqrt{6}\pi m_W/\alpha_W$ is of the order of the
minimum barrier height Eq. (\ref{ewsp}). The so-called ``holy-grail''
function $F$ in the exponent  is
known only in a low energy expansion whose first few terms
are given by \cite{hgew}
\begin{equation}
\label{ewholy}
F^{\rm QFD}(\epsilon ) = 1 -{9\over 8}\ \epsilon^{4/3}
+{9\over 16} \ \epsilon^2 + {\cal O}(\epsilon^{8/3}[1+\ln\epsilon ] ) ,
\end{equation}
where $\epsilon =\sqrt{ s^\prime }/M_0$.  The effective instanton
size entering the running coupling in (\ref{totew}) scales
like $m_W^{-1}$,
\begin{equation}
\rho_\ast=\frac{1}{m_W}\left[
\sqrt{3\over 2}\epsilon^{2/3}+\cdots \right].
\label{ewrho}
\end{equation}
Note, that the first term in the series expression
for the holy grail function, Eq.~(\ref{ewholy}),
corresponds to the ``naive'' tunnelling factor,
Eq.~(\ref{tunnelling}).
Apparently, the total cross section is exponentially
growing for $(m_W\ \ll )\ \sqrt{ s^\prime }\ \ll \ M_0$, but still small
within the region of validity of expansion (\ref{ewholy}).
As anticipated above, in this energy region, it is dominated by the associated
production of a large number of $W$ and $Z$ bosons,
\begin{equation}
\label{wmult}
\langle\ n_W\ \rangle = {\pi\over \alpha_W}\ \biggl[
{3\over 2}\ \epsilon^{4/3} + {\cal O}(\epsilon^2 ) \biggr] .
\end{equation}
 Unfortunately, nothing is known
about the behaviour of the holy grail function for $\sqrt{s^\prime}$
around or above the barrier energy $M_{\rm barrier}^{\rm QFD}$.
 The different terms in the perturbative expansion of
$F^{\rm QFD}$ become comparable in size, and the perturbative
expansion breaks down, just in this most interesting region.
 Unitarity and other arguments along with various assumptions
have been used to argue
\cite{uni} that the decrease of the holy grail function may well
level off at values of order
$F^{\rm QFD}\simeq 1/2$, leading to unobservably small cross sections
of electroweak $B+L$ violation. However, this question is not finally settled.

It is very remarkable that the contribution of QCD-instantons to
deep inelastic scattering strongly ressembles
Eqs. (\ref{totew})-(\ref{wmult}), as first observed for
$g^\ast g$ scattering in Refs. \cite{kr,bbgg} and elaborated
for $\gamma^\ast g$ scattering in Ref. \cite{bb}.
In the Bjorken limit (c.\,f. Fig.~\ref{f2} for the kinematics),
\begin{equation}
\left. \begin{array}{c}
Q^{\prime \,2}=-q^{\prime \,2},\\
\mbox{c.\,m. ${\rm energy}^2$ of $I$-subprocess, } s^\prime =(q^\prime +p)^2\\
\end{array} \right\}\rightarrow \infty\ ;\
x^\prime=\frac{Q^{\prime\,2}}{2pq^\prime}
\ \mbox{ fixed},
\end{equation}
the total subprocess
cross section for instanton-induced chirality violation is found
to have the following structure
\begin{equation}
\label{totQCD}
 \sigma^{(I)\,{\rm tot}}_{\rm QCD} \propto
\exp\left[ -{4\pi\over \alpha_s(\rho_\ast^{-1})}\ F^{\rm QCD} (x^\prime )
 \right] ,
\end{equation}
where \cite{kr,bbgg,bb}
\begin{eqnarray}
F^{\rm QCD}( x^\prime ) &=& 1 -{3\over 2}\
\left( {1-x^\prime\over 1+x^\prime }\right)^2 +
  {\cal O}\left( \left( {1-x^\prime\over 1+x^\prime }\right)^4
\left[ 1 + \ln \left( {1-x^\prime\over 1+x^\prime }\right) \right]
\right) ,
\nonumber\\[3mm]
\rho_\ast&=&\frac{4 \pi}{\alpha_s(Q^\prime) Q^\prime}\left[
3\biggl( {1-x^\prime\over 1+x^\prime }\biggr)^2+\cdots \right].
\label{QCDholy}
\end{eqnarray}
The effective instanton size $\rho_\ast$ in
Eqs.~(\ref{totQCD},\,\ref{QCDholy}), acting
as the characteristic linear dimension,
as well as the scaling form of $F^{\rm QCD}(x^\prime )$
are in accordance with the
substitution rule (\ref{subst}) from QFD to QCD.
The average gluon multiplicity is found to be
\cite{rs},
\begin{equation}
\label{gmult}
\langle\ n_g\ \rangle = {\pi\over \alpha_s}\ \left[ \ 6
\left( {1-x^\prime\over 1+x^\prime }\right)^2 +
 {\cal O}\left( \left( {1-x^\prime\over 1+x^\prime }\right)^3 \right)
\right] .
\end{equation}

It is exactly this similarity between
QFD and QCD in\-stan\-ton-in\-duced scattering processes
which makes the study of the
latter at HERA so interesting.

\vspace{0.2cm}
\section{Instanton-Induced Contributions
to Structure Functions}
In this Section, let us sketch the essential steps in the
pioneering calculation of QCD-instanton
contributions to the (nucleon) structure functions in Ref.~\cite{bb}.
Along the way, we shall emphasize the basic ingredients
as well as the inherent limitations. Finally, a state of the
art evaluation of the $I$-induced contribution to the
$F_2$ structure function of the proton will be presented.

First of all, it is argued \cite{bb} that the celebrated
factorization theorem remains valid
beyond conventional perturbation theory and allows to express
the instanton contribution to the nucleon
structure functions $F_{1,\,2}$ in the familiar form
\begin{equation}
\label{strucfunc}
F_i^{(I)}(x_{\rm Bj},Q^2)=
a_i (x_{\rm Bj} ) \sum_{p=g,q,\bar{q}}
\int_{x_{\rm Bj}}^1 {dx\over x}\ p \left( {x_{\rm Bj}\over x},\mu \right)\
{\cal F}_i^{(I)\,p} \left(x,{Q^2\over \mu^2},\alpha_s (\mu^2 ) \right) .
\end{equation}
In Eq.~(\ref{strucfunc}),
 $a_1=1/2$, $a_2(x_{\rm Bj})=x_{\rm Bj}$, $x$
is the Bjorken variable of the $\gamma^\ast$-parton subprocess ,
 and $\mu$ is the
factorization scale
separating ``hard'' and ``soft'' contributions to the cross section.
The distributions $p(z,\mu )$ of partons $p$ in the nucleon absorb
all information about the dynamics at large distances
and, as usual,  are to be taken from experiment.
By virtue of Eq.~(\ref{strucfunc}),
the theoretical efforts in Ref.~\cite{bb} concentrate
on calculating the ``parton" structure functions
${\cal F}_i^{(I)\,p} \left(x,{Q^2\over \mu^2},\alpha_s (\mu^2 ) \right)$
in the instanton background.
For a  detailed discussion on the familiar and important problem of
infrared (IR) divergencies (associated with integrations over the
instanton size), we have to refer to Ref.~\cite{bb}. In summary, it is
claimed that these divergencies  may be consistently absorbed into
the parton distributions $p(z,\,\mu )$, and an unambiguous,
IR-protected contribution from small instantons may be isolated.

\begin{figure}
\vspace{-0.4cm}
\begin{center}
\epsfig{file=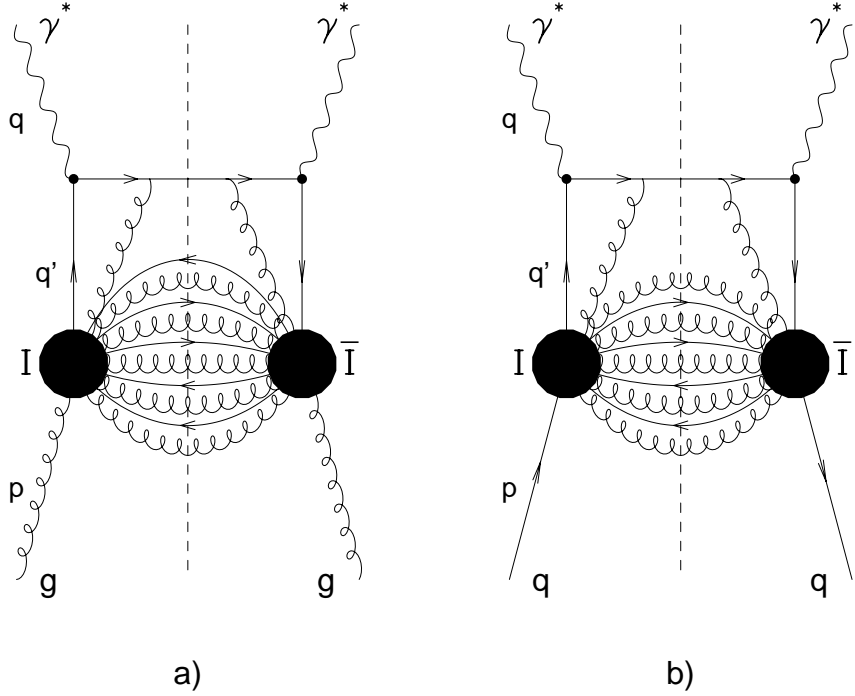,bbllx=58pt,%
bblly=75pt,bburx=534pt,bbury=683pt,width=9cm,angle=90}
\caption[dum]{\label{f2}{\capfont
The contribution of an $\iai$ pair to
to the structure function of a gluon (a) and of a quark (b).
Solid lines are quark zero modes in the case that they are ending
at the instanton (anti-instanton), and quark propagators
in the $I\overline{I}$ background otherwise.
Note that the black blobs denoted by $I(\ai )$ are often
referred to in the text as the $I(\ai )$-subprocess.
}}
\end{center}
\end{figure}

According to the optical theorem
 the parton structure functions
${\cal F}_i^{(I)\,p}$ are related to
the imaginary part of the forward virtual photon-parton matrix element
(c.f. Fig.~\ref{f2})
\begin{equation}
T_{\mu\nu}^{\rm parton}
 = \ii\int d^4z\,e^{\ii qz} \langle\, {\rm parton}(p),\,\lambda\, |
T\{ j_\mu(z) j_\nu (0)\}|\,{\rm parton}(p),\,\lambda\,\rangle .
\label{matrix}
\end{equation}

The calculation of the instanton-induced
contribution to the parton structure
functions then involves the following steps:
\begin{itemize}
\item The path integral expression for the
matrix element (\ref{matrix}) in Euclidean space is expanded
about the instanton/anti-instanton pair configuration, defined via the
so-called valley method \cite{kr,vm}\footnote{
For any fixed values of the
collective coordinates $\{\tau\}$, the pair configuration
($\iai$ ``valley'') is required to  minimize the action
within the subspace orthogonal to $\partial G^{\iai} /\partial \tau_i$.}.
\item Next, the integrations over the large number of collective coordinates
associated with the $\iai$ configuration have to be performed.
\item After Fourier transformation,
the last step consists in rotating the result to Minkowski space and,
thereafter, taking the imaginary part.
\end{itemize}

After a long and tedious calculation, heavily
exploiting the light-cone approximation, Balitsky and Braun
\cite{bb}  succeeded in performing these steps.
Their final answer  for the instanton-induced contribution
to the gluon and quark structure functions,
 derived in the Bjorken limit (c.\,f. Fig.~\ref{f2}),
\begin{equation}
\left. \begin{array}{c}
Q^2=-q^2,\\
\mbox{$\gamma^\ast$--parton c.\,m. ${\rm energy}^2$, } \hat s =(q+p)^2\\
\end{array} \right\}\rightarrow \infty\ ;\  x=\frac{Q^2}{2pq}
\ \mbox{ fixed},
\end{equation}
reads:
\begin{eqnarray}
{\cal F}_{1,\,2}^{(I)\,g}(x,Q^2)
  &\simeq &\sum_q e^2_q
\frac{1}{9(1- x)^2}
\frac{ d^2\pi^{9/2}}{bS(\xi_\ast)[bS(\xi_\ast)-1]}
\left(\frac{16}{\xi_\ast^3}\right)^{n_f-3}
\nonumber\\
&&\mbox{}\times
\left(\frac{2\pi}{\alpha_s(\rho_\ast^{-1})}\right)^{19/2}
\!\!\exp\left[ -
\left(
\frac{4\pi}{\alpha_s(\rho_\ast^{-1})} +2b\right)
S(\xi_\ast)\right] ,
\label{answer}\\[5mm]
{\cal F}_{1,\,2}^{(I)\,q}(x,Q^2)
&\simeq &
\left[\sum_{q'\neq q} e^2_{q'} +\frac{1}{2}e^2_q\right]
 \frac{128}{81(1-x)^3}
\frac{d^2\pi^{9/2} }{bS(\xi_\ast)[bS(\xi_\ast)-1]}
\left(\frac{16}{\xi_\ast^3}\right)^{n_f-3}
\nonumber\\
&& \mbox{}\times
\left(\frac{2\pi}{\alpha_s(\rho_\ast^{-1})}\right)^{15/2}
\exp\left[ -
\left(
\frac{4\pi}{\alpha_s(\rho_\ast^{-1})} +2b\right)
S(\xi_\ast)\right] ,
\label{qanswer}
\end{eqnarray}
where $e_q$ are the electric charges of the quarks,
 $b=11-(2/3)n_f$, and
$d\simeq 0.00363$ (for $n_f=3$ massless flavours) is a constant
which enters the expression for the instanton density \cite{th}.

The classical action of the instanton/anti-instanton pair, $S(\xi )$,
is the most important ingredient in Eqs. (\ref{answer},\,\ref{qanswer}),
since it enters in the exponent.    Due to conformal invariance,
it only depends on the following combination \cite{kr,vm}
of collective coordinates,
\begin{equation}
  \label{xi}
  \xi ={R^2 + \rho_\i^2 +\rho_\ai^2\over \rho_\i\rho_\ai } \, ,
\end{equation}
with $R^2=(x_I-x_\ai )^2$ being the
instanton -- anti-instanton separation, and
$\rho_I$,\,$\rho_\ai$ their sizes, respectively.
For large $\xi$, the action ressembles a ``dipole'' form \cite{cdg,kr,vm}
\begin{equation}
\label{dipaction}
S(\xi ) = 1 -{6\over \xi^2} + {\cal O}({\rm ln}(\xi )/\xi^4 ) .
\end{equation}
Finally, in this ``dipole'' approximation, the effective conformal
parameter $\xi_\ast$ and instanton size $\rho_\ast$ entering
Eqs.~(\ref{answer},\,\ref{qanswer}), read
\begin{eqnarray}
      \xi_\ast & \simeq &
2\  \frac {1+x}{1-x} ,
\label{xisp}\\
     \rho_\ast & \simeq &
\frac{4\pi}{\alpha_s(Q)Q}
     \frac{12}{\xi_\ast^2} .
\label{rhosp}
\end{eqnarray}
At this point a number of important remarks should be made.

Despite the complications associated with the $\gamma^\ast$-parton
dynamics (c.f. Fig.~\ref{f2}), the gluon and quark structure functions
(\ref{answer},\,\ref{qanswer})
apparently exhibit the typical signatures
of an individual $I$-subprocess  cross section
(\ref{totQCD},\,\ref{QCDholy}). Of course, the $I$-subprocess
variables $x^\prime ,Q^\prime$  appearing in
Eqs.~(\ref{totQCD},\,\ref{QCDholy}) are integrated over here
(c.f. also Fig.~\ref{f2})
and effectively substituted by the appropriate
$\gamma^\ast$-parton variables $x,Q$.
Let us note, in particular, that the approximate expressions
(\ref{xisp},\,\ref{rhosp}) agree with the
solutions of the ``saddle-point equations'' -- associated with the
integrations over the collective coordinates --
in case of an individual $I$-subprocess \cite{kr,bbgg}.

The applicability of Eqs.~(\ref{answer},\,\ref{qanswer}) is restricted
to sufficiently large $x$ (c.\,f. Eq.~(\ref{xisp})), since their
derivation  was based
on the large $\xi$ (``dipole'') approximation (\ref{dipaction}) for the
action. A further technical requirement is
$1-x\gg \sqrt{\alpha_s(\rho_\ast^{-1})}$, excluding the
neighbourhood of $x= 1$.

In general, the $\iai$ interaction, $U_{\rm int}(\xi_\ast )
=S(\xi_\ast )-1$, describes the emission and absorption of
gluons from the instanton to the anti-instanton and vice versa
(wavy lines between instanton and anti-in\-stan\-ton in Fig.~\ref{f2}).
It generates  via the Cutkovsky rules  all final state
tree-graph corrections to the leading semi-classical result
(for a formal proof, see Ref.~\cite{cutkov}).  These
final state corrections
are well known to exponentiate \cite{holy}. However,
it has been argued that some initial state
and initial state -- final state corrections exponentiate as well
\cite{muell} and might give rise to additional corrections of order
$4\pi/\alpha_s {\cal O}((1-x)^5)$ in the exponent.

The pre-exponential factor in Eqs.~(\ref{answer},\,\ref{qanswer})
is calculated  only to leading accuracy
in the strong coupling and up to corrections of order ${\cal O}(1-x)$.
This is largely due to the fact, that $I$ -- $\ai$
interactions have been essentially neglected
in the prefactor, unlike the $\iai$  action in the exponent.
\begin{itemize}
\item On the one hand, this refers to the treatment of the
``current quark'' propagating
in the $\iai$ background (c.\,f. Fig.~\ref{f2}).
Its presence gives rise to
great technical complications and, correspondingly,
the results (\ref{answer},\,\ref{qanswer})
 only account for the first nontrivial terms in the
cluster expansion \cite{cluster} of the full current quark propagator
in terms of the known propagator \cite{brown}
in the  background of a single (anti-)instanton.
\item On the other hand, this refers to
the evaluation of the functional determinants
entering the pre-exponential factor. An improvement
based on the $\iai$ valley is under way \cite{frs}.
\end{itemize}

\begin{figure}
\vspace{-0.4cm}
\begin{center}
\epsfig{file=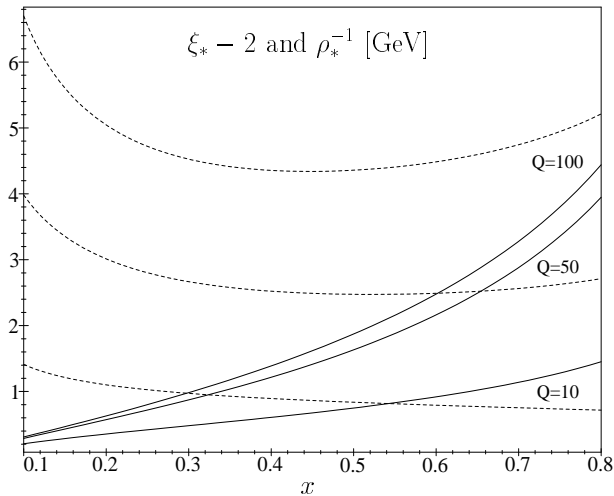,bbllx=124pt,%
bblly=282pt,bburx=491pt,bbury=584pt,width=8cm}
\caption[dum]{\label{f3}{ \capfont
The effective conformal parameter $\xi_\ast-2$ (solid) and
inverse instanton size $\rho_\ast^{-1}$  (dashed),
obtained as solutions
of the saddle point equations, Eq.~(\ref{saddleeq}),
for a range  of $Q$ values (in GeV) and $n_f=3$.
 }}
\end{center}
\end{figure}
Despite these considerable formal restrictions in the derivation of
the gluon and quark structure functions (\ref{answer},\,\ref{qanswer}),
it is tempting to try and evaluate these results within an
experimentally accessible regime of $x_{\rm Bj}$ and $Q^2$.
In order to hopefully enlarge
the kinematical region where  Eqs.~(\ref{answer},\,\ref{qanswer})
may be qualitatively trusted, we have heuristically applied the
following ``improvement'' steps \cite{rs} (see also Ref.~\cite{bb}):
\begin{itemize}
\item Throughout, in Eqs.~(\ref{answer},\,\ref{qanswer}), we use the
$\iai$ valley action $S(\xi)$ of Refs.~\cite{kr,vm}, rather than its ``dipole''
approximation (\ref{dipaction}). This action represents an
extension of Eq.~(\ref{dipaction}) to arbitray values of $\xi \ge 2$.
It is worth pointing out that it continuously interpolates between
$S = 1$ (the sum of individual instanton and anti-instanton actions)
at large $\xi$, and $S = 0$ for $\xi = 2$. These
limiting situations correspond to a widely separated, non-interacting
$\iai$-pair  for $\xi\to \infty$ and the perturbative configuration of a
collapsing and annihilating  $\iai$-pair for $\xi \to 2$
(i.e. for $R\to 0,\rho_I=\rho_\ai$).
\item We replace the approximate
expressions (\ref{xisp},\,\ref{rhosp}) for the effective
conformal parameter $\xi_\ast$ and instanton size $\rho_\ast$,
respectively, by the solutions of the exact ``saddle-point''
equations \cite{kr,bbgg,bb},
\begin{eqnarray}
\sqrt{\hat s} \rho_\ast  &=& \frac{8\pi}{\alpha_s( \rho_\ast^{-1})}
\sqrt{\xi_\ast-2}\,S'(\xi_\ast)\,,
 \label{saddleeq}      \\
Q\rho_\ast &=&
\frac{4\pi}{\alpha_s(\rho_\ast^{-1})}(\xi_\ast-2)S'(\xi_\ast)
-\rho_\ast {\partial \over \partial\rho_\ast}\biggl(
{2\pi\over \alpha_s(\rho_\ast^{-1})} \biggr) S(\xi_\ast ) \, ,
\nonumber
\end{eqnarray}
where $S'(\xi)$ is the derivative of
the valley action $S(\xi)$ with respect to $\xi$.
We have numerically solved Eqs.~(\ref{saddleeq})
(see Fig.~\ref{f3}),
using the two-loop expression for the running coupling
$\alpha_s(\rho_\ast^{-1})$ with three active flavors,
and the value $\Lambda_{\overline{\rm MS}}^{(3)} = 365$ MeV.  It
corresponds to $\alpha_s (m_\tau)= 0.33$ \cite{ALEPH}.
\end{itemize}

\begin{figure}
\vspace{-0.4cm}
\begin{center}
\epsfig{file=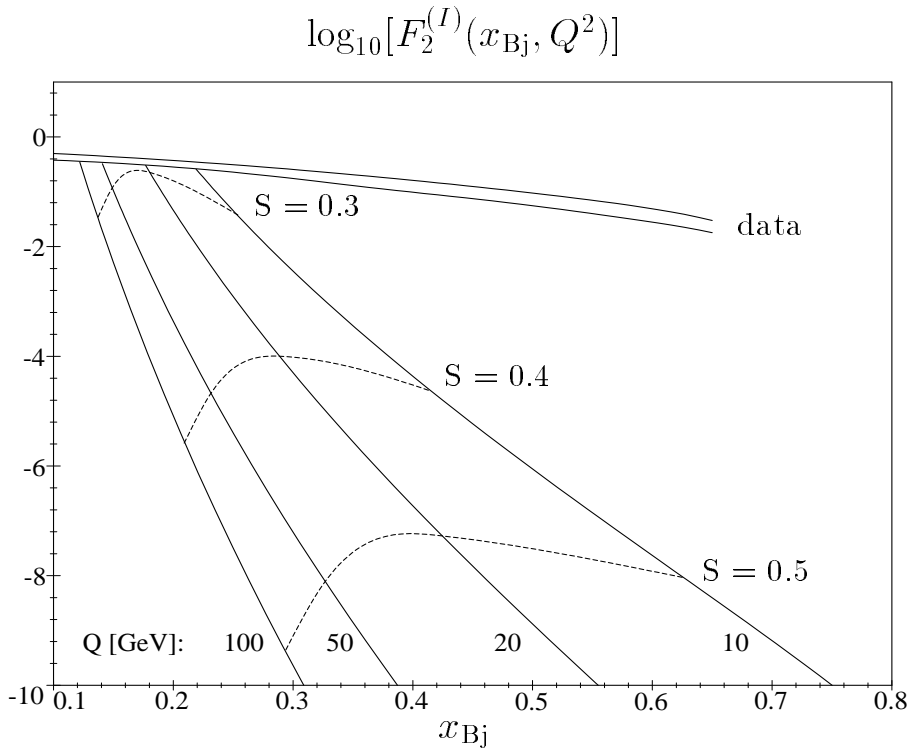,bbllx=113pt,bblly=277pt,%
bburx=491pt,bbury=617pt,width=12cm,height=10cm}
\caption[dum]{ \label{f4}{\capfont
The logarithm of the instanton-induced contribution to
the structure function $F_2$ of the proton,
$\log_{10} [ F^{(I)}_2 (x_{\rm Bj},Q^2) ]$, Eq. (\ref{strucfunc}), as
a function of $x_{\rm Bj}$, for a range of $Q$ values (in GeV) and $n_f=3$.
The curves denoted by ``data'' are to roughly represent
the trend of the experimental data for $F_2$ within
the same range of $Q$.
The dashed curves correspond to constant values of the valley action,
$S(\xi_\ast (x_{\rm Bj},Q^2 ) )$.
}}
\end{center}
\end{figure}
The $I$-contribution to the nucleon structure function
was finally obtained from convoluting the ``improved''
Eqs.~(\ref{answer},\,\ref{qanswer})
with very simple phenomenological expressions
\cite{gs} for the gluon, $u$ quark
and $d$ quark distributions, $g(z)=(3/z)(1-z)^5$,
$u(z)=(2/\sqrt{z})(1-z)^3$,  $d(z)=(1/\sqrt{z})(1-z)^3$, respectively.
In view of the qualitative nature of this study,
they turn out to be quite adequate for a factorization
scale $\mu \simeq \rho_\ast^{-1}$, which is natural in this context.
It turns out that over
the whole $x_{\rm Bj}$ range considered the $\gamma^\ast g$ contribution to
$F_2^{(I)}$ dominates. The sea-quark contributions can be
neglected throughout the $x_{\rm Bj}$ range considered.

The resulting instanton-induced contribution to the structure function
$F_2$ of the proton is displayed in Fig.~\ref{f4}.
The expected very strong rise of the $I$-induced contribution with
decreasing $x_{\rm Bj}$ is both apparent and suggestive!

Unfortunately, any further conclusions
directly reflect the ($x_{\rm Bj},Q^2$) region where
the above approximations are supposed to hold.
For instance, the dashed lines in Fig.~\ref{f4} define the
boundaries of various ``fiducial'' regions corresponding to
values of  $S(\xi_\ast (x_{\rm Bj},Q^2 ) ) \geq 0.5,\,0.4,\,0.3$.
As mentioned before, some authors \cite{uni}
have advocated $F^{\rm QCD}_{\rm min} =S(\xi_\ast )_{\rm min}=1/2$
as a saturating value for the holy grail function $F^{\rm QCD}$.
The minimum value of $Q$ considered in Fig.~\ref{f4} is determined by
the requirement that the effective instanton size should be
sufficiently small.
At $Q=10$ GeV one finds $\rho_\ast \simeq 1$ GeV$^{-1}$  (c.\,f.
Fig.~\ref{f3}).

\vspace{0.2cm}
\section{Phenomenology of Instanton Induced Particle Production
at HERA}
There are three main reasons which favour  experimental
sear\-ches for in\-stan\-ton-induced ``footprints'' in the
multi-particle final state over searches via the
 structure functions, being the
most inclusive observables in deep inelastic scattering.
\begin{itemize}
\item On the one hand, the only experimental signal
      for QCD-instantons in the structure functions
      could be in form of an excess over the expected
      inclusive leptoproduction rate.
      However, enhancements at small $x_{\rm Bj}$ are also expected
      from other competing mechanisms
      like ``(perturbative) Reggeization''. Therefore,
      the structure functions are only of limited value in searches
      for manifestations of QCD-instantons.
\item On the other hand, as we shall see, the instanton-induced
      {\it final state} is distinguished by a quite spectacular
      event topology together with a characteristic flow of
      flavour quantum numbers.
\item Furthermore, the additional possibility of imposing experimental
      cuts  on kinematical variables of the final state
      may well allow to restrict the $I$-subprocess variables
      $x^\prime ,Q^\prime ,\ldots$ within a theoretically controllable
      regime despite small $x_{\rm Bj}$.
      Along these lines one may hope to bridge the
      substantial gap between
      the regime of larger $x^\prime \gwig {\cal O}(0.1)$, where the
      $I$-subprocess cross sections may be theoretically estimated, and
      the small $x_{\rm Bj}$ regime, $x_{\rm Bj} \lwig O(10^{-3})$,
      where the bulk of HERA data is accumulating at present.
\end{itemize}
To elaborate on the last two aspects is the purpose of this Section.

\bigskip
\begin{figure}
\begin{center}
\epsfig{file=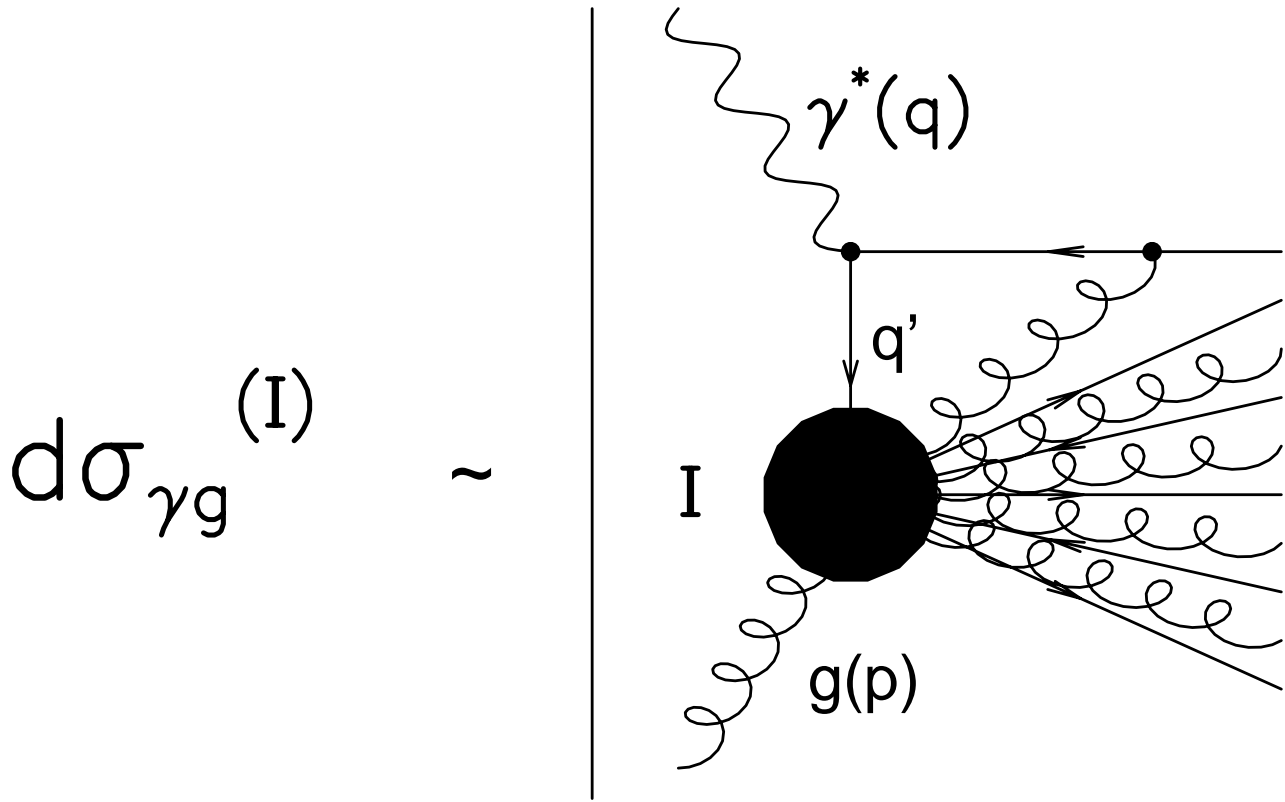,bbllx=170pt,%
bblly=107pt,bburx=465pt,bbury=708pt,width=11cm,angle=90}
\caption[dum]{\label{f5}{ \capfont
Graphical display of the  instanton-induced contribution
to the cross section of $\gamma^\ast g$ scattering for $\Delta Q^5 =2n_f=6$.
 }}
\end{center}
\end{figure}

A graphical display of the modulus squared of the relevant $\gamma^\ast g$
matrix element,  along with
various four-momenta of interest,
is presented in Fig.~\ref{f5}.
Its structure suggests
that the instanton-induced contribution
to the differential cross sections for
$\gamma^\ast g(q)\rightarrow \cdots$ can be
written in a ``canonical'' convolution form, as familiar from perturbative
QCD:
\begin{equation}
\label{convolution}
d\sigma^{(I)}_{\gamma^\ast p} (x,Q^2,\ldots )
\sim \sum_{p^\prime}
\int^{Q^2/x}  {dQ^{\prime 2}\over Q^{\prime 2}}
\int_x^1 {dx^\prime\over x^\prime}
f^{(I)}_{\gamma^\ast p^\prime}
\left( {x\over x^\prime},Q^{\prime 2} \right)
d\sigma^{(I)}_{p^\prime p}
(x^\prime , Q^{\prime 2},\ldots ).
\end{equation}
The integrations in Eq.~(\ref{convolution}) extend over the
variables $x^\prime,\,Q^{\prime \, 2}$, referring as in Sect.~2 to the
instanton-induced subprocess (denoted by $I$ in
Fig.~\ref{f5}),
\begin{equation}
\begin{array}{lcl}
Q^{\prime 2}&=&-q^{\prime 2}\\[1.5mm]
s^\prime&=&(q^\prime+p)^2\\
\end{array}
\hspace{1cm}x^\prime=\frac{Q^{\prime 2}}{2pq^\prime}
        =\frac{Q^{\prime 2}}{s^\prime+Q^{\prime 2}}.
\end{equation}
Their definition is completely analogous to the standard $e^\pm N$
variables $Q^2=-q^2$ and $x_{\rm Bj}=Q^2/2Pq$
referring to the nucleon target of momentum $P$.
The parton (gluon) momentum fraction with respect to the proton
is $z=x_{\rm Bj}/x =(p q)/(P q)$ and
\begin{equation}
0<x_{\rm Bj}\leq x \leq x^\prime \leq 1.
\label{ordering}
\end{equation}

The conditions for the validity of Eq.~(\ref{convolution}) (beyond the
set of approximations inherent in Ref.~\cite{bb}), as well as a
determination of the ``splitting function'' $f^{(I)}_{\gamma^\ast p^\prime}
(  z^\prime,Q^{\prime 2} )$, associated with the
propagation of the current quark
in the instanton background, are presently under active
investigation \cite{mrs}. To establish a structure of type (\ref{convolution})
is quite an important task both from a theoretical point of view and also
for further studies of instanton-induced
phenomenology by
means of Monte Carlo methods \cite{grs}. For the time being, we shall
simply assume that Eq.\,(\ref{convolution}) is valid approximately.

It is quite plausible that the
``splitting function'' $f^{(I)}_{\gamma^\ast p^\prime}
( z^\prime,Q^{\prime 2} )$ in Eq.~(\ref{convolution})
only exhibits a relatively mild dependence on its variables.
In contrast, the $I$-subprocess cross sections
$d\sigma^{(I)}_{p^\prime p} (x^\prime , Q^{\prime 2},\ldots )$ bring
in the main dependences (of exponential type) and, of course,
are most interesting from the physics point of view.
Accordingly, we have concentrated our theoretical efforts in
Ref.~\cite{rs} on calculating the crucial observables characterizing
the $I$-subprocess, such as normalized $1,2,\ldots$-parton
inclusive cross sections,
\begin{equation}
\frac{1}{\sigma_{p^\prime p}^{(I)\,\rm tot}(x^\prime , Q^{\prime 2})}
d\sigma^{(I)}_{p^\prime p} (x^\prime , Q^{\prime 2},\ldots ) ,
\end{equation}
along with the respective average parton multiplicities,
transverse momentum (flow), etc.
Corresponding to the restrictions discussed in Sect.~3 in the context
of the structure functions, the $I$-subprocess
variables should not be too small,
$x^\prime \gwig {\cal O}(0.1)$, $Q^\prime \gwig {\cal O}(10$ GeV), say.

Since a calculation of the ``splitting function''  in
Eq.~(\ref{convolution}) is still under way \cite{mrs}, a
discussion of expected event rates has to be deferred
to a later stage. In the present analysis,
we only make use of information abstracted from our calculations
of the $I$-subprocess observables \cite{rs} along with
HERA kinematics. This is sufficient, however, to obtain
important insight into the expected event topology in the most
interesting regime of small $x_{\rm Bj}$. Moreover,  the connection
between kinematical quantities measurable in the laboratory system and the
variables controlling the instanton subprocess may be studied.

Specifically, we use the following set of working hypotheses about the
$I$-subprocess:

\noindent
{\it i) Isotropy}: In its c.m. system, $\vec{q^\prime}+\vec{p}=0$, the
instanton-induced multi-parton production is supposed to proceed
{\it isotropically}. We may imagine a  ``fireball'' in $S$-wave
configuration, decaying into gluons and at least $2n_f-1$ quarks,
including {\it strangeness}~(!) and possibly charm, if kinematically
allowed (c.\,f. Fig.~\ref{f5}).

\noindent
{\it ii) Dependence on $x^\prime,\ Q^{\prime\,2}$}: The
$I$-subprocess cross sections $d\sigma^{(I)}_{p^\prime p}
(x^\prime , Q^{\prime 2},\ldots )$ are expected to strongly
decrease with increasing $Q^{\prime 2}$ for fixed $x^\prime$ and to
strongly increase with decreasing $x^\prime$ for fixed $Q^{\prime 2}$
(c.\,f. Sect.~2).
As discussed above and in Sect.~3, it remains uncertain, however,
how long the cross sections continue to increase towards
$x^\prime \rightarrow 0$.

\begin{figure}
\vspace{-0.4cm}
\begin{center}
\epsfig{file=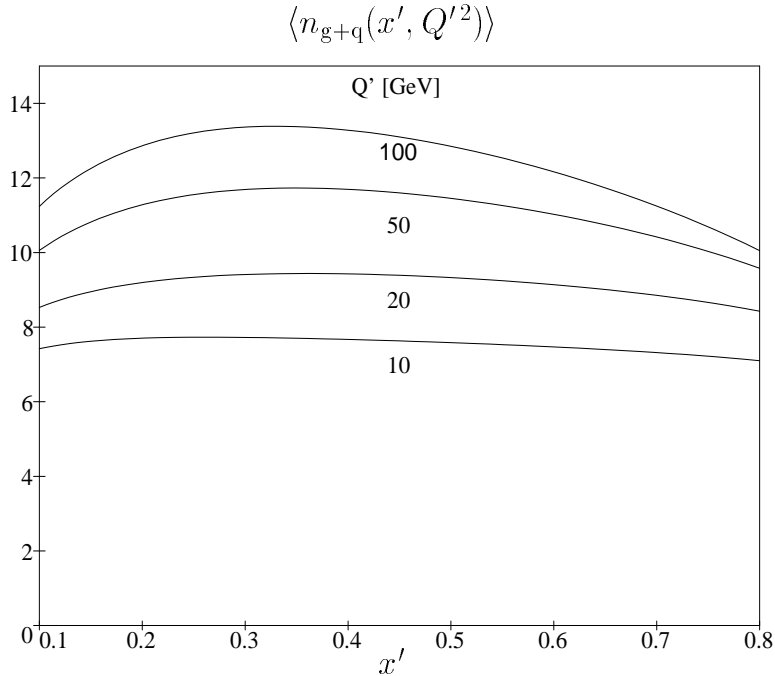,bbllx=123pt,%
bblly=282pt,bburx=491pt,bbury=612pt,height=9cm}
\caption[dum]{\label{f6}{ \capfont
The average, total parton multiplicity associated with the
$I$-sub\-pro\-cess as a function of $x^\prime$, for different
values of $Q^\prime$ and $n_f=3$, from Ref.~\cite{rs}.
 }}
\end{center}
\end{figure}

\noindent
{\it iii) Multiplicity}: The total multiplicity
associated with the $I$-subprocess is expected to be
large,
\begin{equation}
\langle n_{\rm g + q}(x^\prime,Q^{\prime 2})\rangle
\sim {\cal O}\left(\frac{\pi}{2\alpha_s}\right)+2n_f-1\gwig
{\cal O}(10),\label{nav}
\end{equation}
on the parton level, leading typically to ${\cal O}(20\div 30)$ particles
after hadronization. In Fig.~\ref{f6}, we display $\langle n_{\rm g + q}
(x^\prime,Q^{\prime 2})\rangle$
as calculated in Ref.~\cite{rs}.

At small values of $x^\prime$ and large $Q^{\prime 2}$, the multiplicity
obtains a large contribution from
gluons and peaks around $x^\prime \approx 0.2\div 0.3$, whereas
at large $x^\prime$ the $2n_f-1=5$ produced quarks dominate.
The peaking of the gluon multiplicity at small, non-vanishing
$x^\prime$ actually has an appealing interpretation:

For large values of $Q^{\prime 2}$ the
coefficient of $\pi/2\alpha_s$ in Eq.\ (\ref{nav}) turns out to be \cite{rs}
$4(\xi_\ast-2)\,S'(\xi_\ast)$, involving the derivative of the valley
action with respect to the conformal parameter $\xi$, taken at the saddle point
value $\xi_\ast(x^\prime,Q^{\prime 2})$ (c.f. Fig.~\ref{f3}). As discussed in
Sect.~3, the full valley action smoothly interpolates between a
non-interacting, infinitely separated instanton/anti-instanton pair
for $\xi \rightarrow \infty$ (probed for $x^\prime \rightarrow 1$) and the
perturbative vacuum for $\xi \rightarrow 2$ (probed for
$x^\prime \rightarrow 0$). Hence, in both limits
a {\it decrease} of the gluon multiplicity matches well the intuition!
The peak of the multiplicity inbetween corresponds to the maximal
variation of the action with $\xi$.

We also note the substantial increase of the gluon multiplicity with
increasing $Q^{\prime 2}$, which at large $Q^{\prime 2}$ mainly reflects
the running of $\alpha_s$ in Eq.\ (\ref{nav}).

\noindent
{\it iv) $k_\bot$ signature}: The transverse momenta of the
partons emerging from the instanton subprocess and the one
of the current-quark jet (c.f. Fig.~\ref{f5}) are expected to be
``semi hard'', typically of order \cite{rs,bb}
\begin{eqnarray}
|k_{\bot\,i}|\sim \frac{<k_{\bot\,{\rm tot}}>}{<n>}
=\frac{\pi}{4}\frac{\sqrt{s^\prime}}{<n>}
&\sim& \alpha_s \frac{\sqrt{s^\prime}}{2};\ i\subset I,
\ \ {\rm and}\\[1mm]
|k_{\bot\,{\rm current\ quark}}|&\sim & \sqrt{\alpha_s Q^2}.
\end{eqnarray}

\bigskip
\noindent
Given this plausible generic input {\it i) - iv)}, we may now ask, how
instanton-induced events
would look like in the H1/ZEUS detectors.
\bigskip

First of all, we observe that as a direct consequence of the {\it isotropy}
assumption {\it i)}, the (pseudo) rapidity distribution of a single final state
parton in the $I$-c.m. system takes the form
\begin{equation}
 \frac{1}{\sigma^{(I\,)\,{\rm tot}}}\frac{{\rm d}\sigma^{(I\,)}
(x^\prime , Q^{\prime 2},\eta_I )}{{\rm d}\eta_I {\rm d}\phi_I}=
\frac{1}{4\pi}\frac{<n (x^\prime,Q^{\prime 2})>}{\cosh(\eta_I)^2},
\end{equation}
i.e. it is strongly peaked in (pseudo) rapidity $\eta_I=
-\ln\tan(\theta_I/2)$ around $\eta_I=0$ with a half width of
\begin{equation}
 \Delta \eta_I \approx\pm 0.9.
\end{equation}
The shape and width of the distribution in pseudo rapidity remains,
of course, very similar in the HERA laboratory system,
for kinematical configurations where the $I$-c.m. system
is dominantly boosted longitudinally. Depending on the values of
the various subprocess variables, the peak position
($\eta_I=0$) fluctuates in general over the available range of
$\eta^{\rm lab}$ (for given $x_{\rm Bj}$ and $y$ viz. $Q^2$):
\begin{eqnarray}
\eta_I^{\rm lab}&=&\frac{1}{2}\ln \left(\frac{E_P}{y E_e}\frac{x_{\rm Bj}}
{x_\gamma}\left[\frac{1-x^\prime}{x}+x_\gamma(1-y)+
(1-x_\gamma)(\frac{x^\prime}{x}-1)\nonumber\right.\right. \\
& &\left.\left.-2\sqrt{(1-y)x_\gamma(1-x_\gamma)(\frac{x^\prime}{x}-1)}
\cos\,\chi\right]\right).
\label{etalab}
\end{eqnarray}
In analogy to the standard $y$ variable, we have introduced in
Eq.~(\ref{etalab})
  the $q^\prime$ momentum fraction
\begin{equation}
0<x_\gamma=\frac{q^\prime p}{q p}<1.
\end{equation}
The variable $\chi$ denotes the azimuthal angle of the vector
$\vec{q^\prime}$ in the HERA laboratory frame. Due to momentum
conservation,
$|\vec{q}_\bot^{\,\prime} |\sin\chi =-k_{y\,{\rm current\ quark}}$,
the component of the current-quark momentum out of the
$ee^\prime P$ scattering plane.

\begin{figure}
\vspace{-0.4cm}
\begin{center}
\epsfig{file=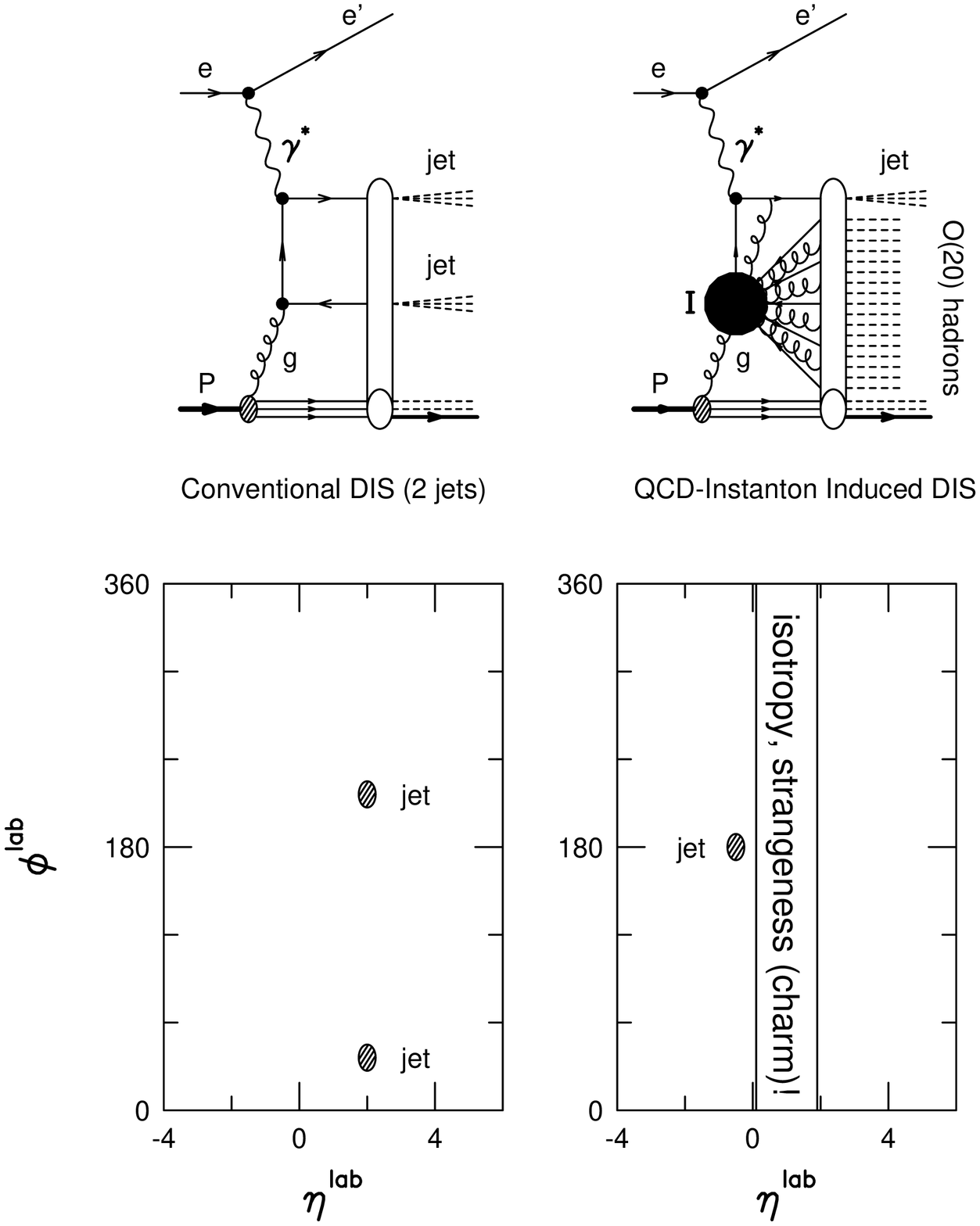,%
bbllx=20pt,bblly=47pt,bburx=552pt,bbury=713pt,width=12cm,angle=0}
\caption[dum]{\label{f7}{ \capfont
The dominant matrix element and $(\eta^{\rm lab},\phi^{\rm lab})$-plot
 of a typical instanton-induced event is
contrasted with the canonical two-jet configuration in perturbative QCD.
 }}
\end{center}
\end{figure}
In Fig.~\ref{f7}, the dominant matrix element and
($\eta^{\rm lab},\phi^{\rm lab}$)-plot
of a typical instan\-ton-induced event is
contrasted with the canonical two-jet configuration in perturbative QCD.

Clearly, the ``0th level'' signature to watch out for is a densely
      populated hadronic ``band'' in the
      $(\eta^{\rm lab},\phi^{\rm lab})$-plane, centered at a fluctuating
      value (\ref{etalab}) of $\eta_I^{\rm lab}$.
      This striking multi-hadron final
      state originates from $8\div 10$ ``semi-hard'' jets
      (c.\,f. Fig.~\ref{f6}),
      always includes strangeness and is characterized by a
      width $\triangle \eta_I^{\rm lab}=\pm 0.9$.
Let us point out two observables, which -- on an event-by-event basis --
appear to be particularly sensitive to this event structure.
\begin{itemize}
\item {\it The (transverse) energy flow},
      ${\rm d} E_{(\bot)}/{\rm d} \eta^{\rm lab}$ (integrated over
      $\phi^{\rm lab}$), will
      exhibit a strong enhancement at the position
      $\eta^{\rm lab}=\eta_I^{\rm lab}$ of
      the ``band'', since each of the $8\div 10$ instanton-induced jets
      contributes a
      comparable energy into a single $\eta^{\rm lab}$ bin of
      width $\approx 1.8$.
      If, in addition, the current-quark jet is isolated from the
      ``band'' (see
       below), one even expects a double-peak structure in
      ${\rm d} E_{(\bot)}/{\rm d} \eta^{\rm lab}$.
      The energy flow signature may well be less affected by hadronization
      than patterns associated with individual tracks.
\item {\it Pseudo sphericity}: The usual event-shape variables like sphericity
      and aplanarity should be useful tools in analysing the manifestations of
      an isotropic instanton-induced subprocess in the final state.
      Of particular
      sensitivity appears the so-called pseudo sphericity \cite{spikes},
      which incorporates only transverse information from the event in terms
      of the azimuthal angles $\phi_i$ of the N final state hadrons:
\begin{equation}
\mbox{Pseudo sphericity}=1-\frac{1}{N}\sqrt{
(\sum_{i=1}^N \cos \phi_i)^2 +
(\sum_{i=1}^N \sin \phi_i)^2} .
\end{equation}
Apparently, it equals to 1 in the c.m.~system of an isotropical event and
vanishes for a single ``collimated'' jet.
\end{itemize}

As a next level of sophistication, we  study the effects of
kinematical cuts on suitable final-state variables
with the aim, to further enhance the event topology and to unfold and/or
restrict the $I$-subprocess variables $x^\prime,\,Q^\prime$.
Of course, this has to be achieved
\begin{itemize}
\item without affecting significantly the expected size of the $I$-subprocess
      cross section ${\rm d} \sigma^{(I)}$
      (by exploring input {\it ii)} above);
\item such that the $I$-subprocess is invoked in a kinematical
      region of ($x^\prime, Q^\prime$), where it induces
      a high  average multiplicity $\langle n_{\rm g+q}\rangle$ according to
      Fig.~\ref{f6}.
\end{itemize}

The key aspect is to focus on an event topology corresponding to an
{\it isolated} (semi-hard) current-quark jet (c.\,f. input {\it iv)}
above: $k_\bot
\approx \sqrt{\alpha_s Q^2}$) {\it in addition to} a hadronic ``band'' in
($\eta^{\rm lab},\phi^{\rm lab}$) as discussed in ``level 0'' above.
To this end, let us consider the constraints on the (internal) subprocess
variables $x_\gamma,\, x^\prime,\,Q^\prime \ldots$,
implied by the following reasonable separation criteria:
\begin{itemize}
\item The hadronic ``band'' should be contained in the (central)
      detector, such that its peak position satisfies
      $|\eta_I^{\rm lab}|\lwig 1$,
      say (in practice, this upper bound for $\eta_I^{\rm lab}$
      may well be pushed up to $\sim 2$).
\item The current-quark jet is required to be separated in
      $\eta^{\rm lab}$ from the hadronic ``band'' (centered around
      $\eta_I^{\rm lab}$) by
\begin{equation}
\triangle\eta \equiv\eta_I^{\rm lab}-
\eta_{\rm current\ quark}^{\rm lab}\ \left\{\begin{array}{l}
\lwig -1.5 \mbox{ or }\\
\gwig +1.5 \\
\end{array}
\right.
\end{equation}
\item A minimal transverse momentum, $k_{\bot\,{\rm current\ quark}}
      \gwig 4$ GeV, is required for the current-quark jet.
\end{itemize}

\begin{figure}
\vspace{-0.4cm}
\begin{center}
\epsfig{file=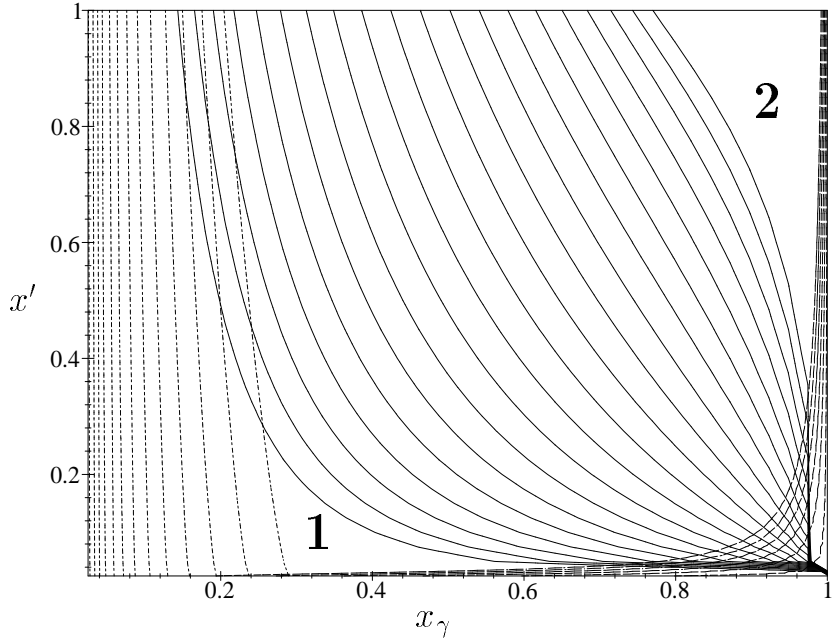,bbllx=100pt,%
bblly=278pt,bburx=493pt,bbury=581pt,width=11cm}
\caption[dum]{\label{f8}{ \capfont
In the displayed regions $1,2$ of the ($x_\gamma,x^\prime$)-plane,
the current-quark jet is isolated from the instanton-induced
``band'' (centered at $\eta_I^{\rm lab}$). The excluded domains are:
$-1.5\le \triangle\eta \equiv\eta_I^{\rm lab}-
\eta_{\rm current\ quark}^{\rm lab}\le 1.5$ (solid lines),
$1\le \eta_I^{\rm lab}$ (short dashes) and
$k_{\bot\ \rm current\ quark}\le 4$~GeV (long dashes).
The parameters are $E_p=820$ GeV, $E_e=30$ GeV, $x_{\rm Bj}=0.001$, $y=0.5$
($Q=7$ GeV),
$x=0.025$ ($\sqrt{\hat{s}}=44$ GeV) and
$\vec{q^\prime}$ azimuthal angle $\chi=0$.
}}
\end{center}
\end{figure}

In Fig.~\ref{f8} we have displayed the resulting restrictions
on the internal subprocess variables
$x_\gamma,\ x^\prime$, for a typical set of
fixed ``external'' parameters $x_{\rm Bj},$$y,$$x$.
Apparently, after imposing the isolation requirements for the
current-quark jet we are left with
two allowed, `triangular' regions 1 and 2,  in the
($x_\gamma,x^\prime$)-plane. The central portion of Fig.~\ref{f8} is
excluded by the cut on $\triangle \eta$ (solid lines), with the left- (right-)
hand boundary corresponding to $\triangle \eta=+1.5\ (-1.5)$. The excluded
portion on the left (short dashes) refers to the hadronic
``band'' being centered within $1 \le \eta_I^{\rm lab}\le 2.2$, with
$\eta_I^{\rm lab}=1$ located on the right. Hence, if a value of
$\eta_I^{\rm lab}$ {\it above} 1 is experimentally tolerable, the allowed
region 1 increases significantly. Finally, the main effect of the requirement
$k_{\bot\ \rm current\ quark}\ge 4$ GeV , is to
set a lower limit to the involved values of $x^\prime$, and
to  exclude the region $x_\gamma$ very close to 1 (long dashes).

These results demonstrate that, indeed, kinematical cuts
of the type considered here, may well restrict the $I$-subprocess
variables $x^\prime ,Q^\prime$ to regions where
the computation \cite{rs}  of $d\sigma_{p^\prime p}^{(I\,)}
(x^\prime ,Q^\prime )$ may be trusted (e.g. within regime 2 of Fig.~\ref{f8}).

\begin{figure}
\vspace{-0.4cm}
\begin{center}
\epsfig{file=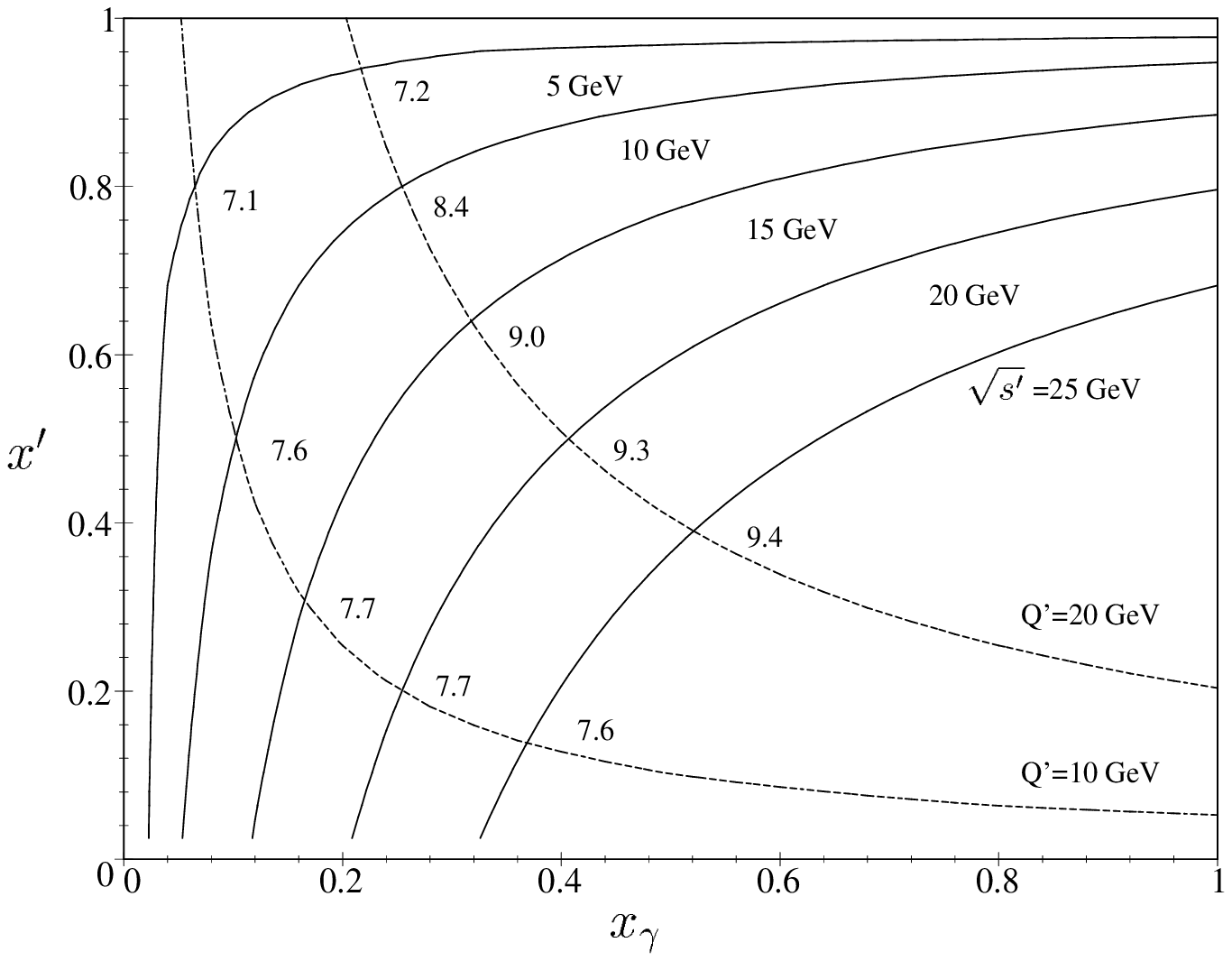,bbllx=100pt,%
bblly=277pt,bburx=491pt,bbury=583pt,width=11cm}
\caption[dum]{\label{f9} {\capfont
Lines of constant invariant mass of the $I$-subprocess, $\sqrt{s^\prime}$,
and constant $Q^\prime$ (dashed), versus the
subprocess variables $x^\prime$ and $x_\gamma$,
for the same parameters as in Fig.~\ref{f8}.
The numbers at the crossing points denote the respective
$I$-subprocess multiplicities $\langle n_{\rm g+q}\rangle$ according to
Fig.~\ref{f6}.
 }}
\end{center}
\end{figure}

According to
our input {\it ii)} above and Fig.~\ref{f9}, regime 1 in Fig.~\ref{f8} will
presumably be associated with considerably higher rates, since it
typically corresponds to much
smaller values of $Q^\prime$ and $x^\prime$ than regime 2. From
Fig.~\ref{f9} we also infer a comfortably high total parton multiplicity
$\langle n_{\rm g+q}\rangle \approx 8$ in regime 1, as well as
an energy/parton $=\sqrt{s^\prime}/8\gwig 3$ GeV in the $I$-c.m. system.


\begin{figure}
\vspace{-0.4cm}
\begin{center}
\epsfig{file=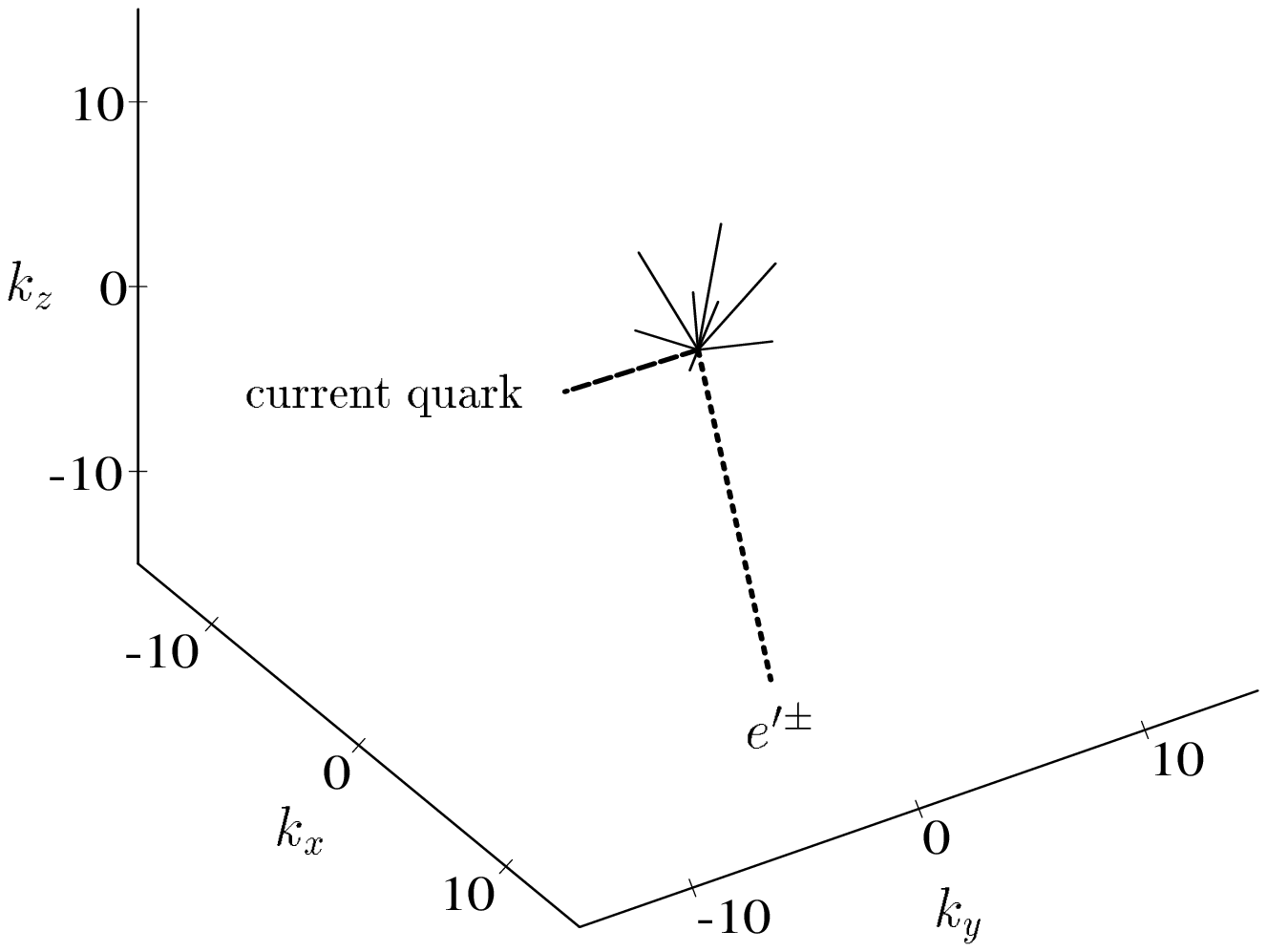,%
bbllx=108pt,bblly=298pt,bburx=488pt,bbury=581pt,width=12cm}
\caption[dum]{\label{f10} {\capfont
3d-momentum display (in GeV) for a typical instanton-induced event before
hadronization,
in the HERA laboratory frame, satisfying the kinematical
cuts discussed in the text. The current-quark jet is well isolated.
Not shown are the incoming proton ($+z$ direction) and $e^\pm$
($-z$ direction), as well as the proton fragments.
The parameters are as in Fig.~\ref{f8} and moreover,
$x_\gamma=0.32,\ x^\prime=0.1$, such that
$Q^\prime\approx 8$ GeV, $\sqrt{s^\prime}\approx 24$ GeV and
$\langle n_{\rm g+q}\rangle \approx 8$.
 }}
\end{center}
\end{figure}

\begin{figure}
\begin{center}
\epsfig{file=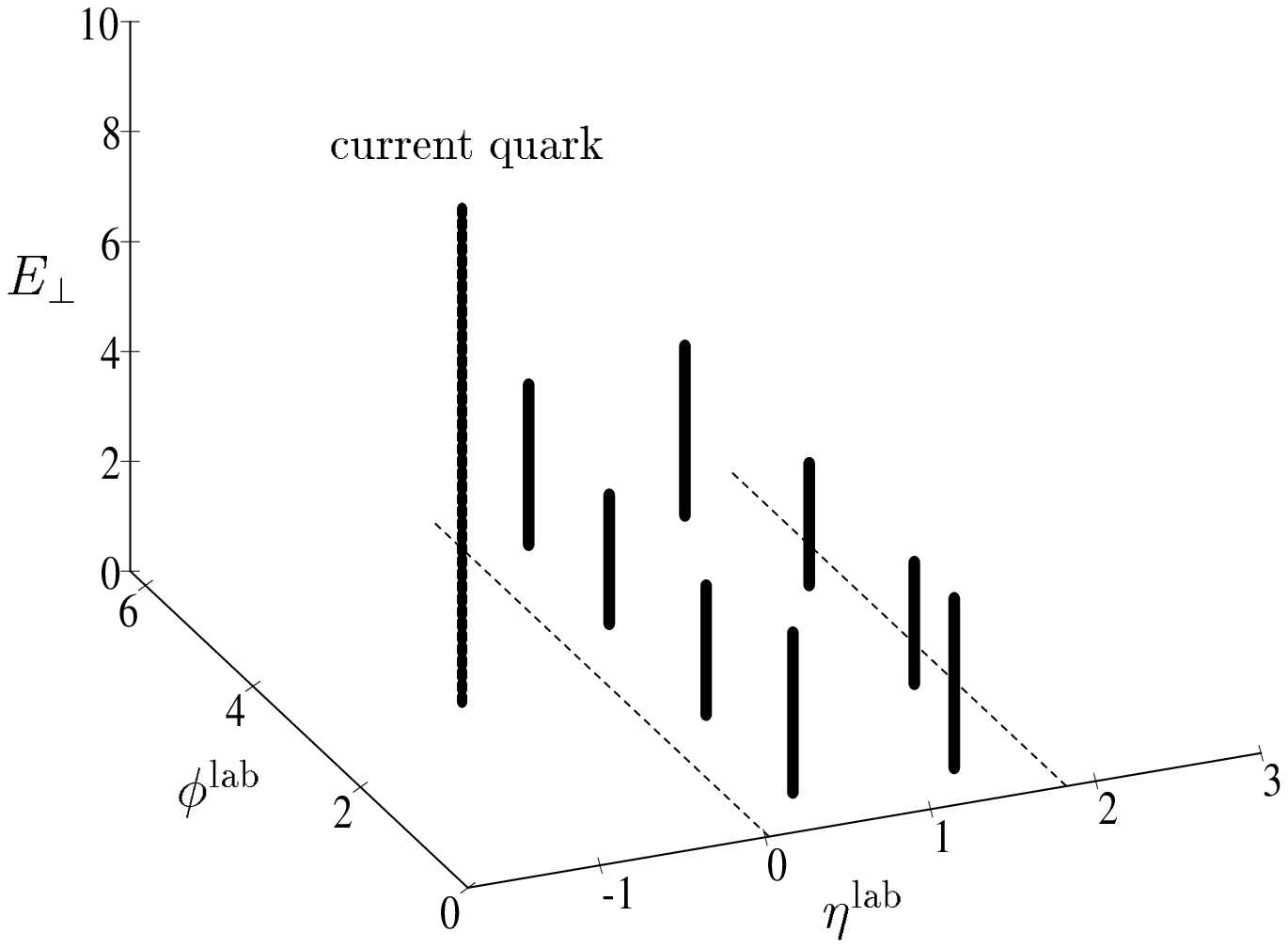,%
bbllx=100pt,bblly=293pt,bburx=492pt,bbury=582pt,width=12cm}
\caption[dum]{\label{f11}{ \capfont
Lego plot of the transverse energy in GeV (before hadronization)
for the same instanton-induced event
as in Fig.~\ref{f10}, satisfying the kinematical cuts discussed in the
text. Not shown are the scattered $e^\pm$ and the proton fragments.
}}
\end{center}
\end{figure}

Let us finally illustrate in
Figs.~\ref{f10},~\ref{f11} an event in the HERA laboratory frame,
corresponding both to a striking signature (isolated
current-quark jet along with a densely populated hadronic ``band'') and
favourable rate/multiplicity conditions for the instanton subprocess.
We note the corresponding event-shape variables (on the parton level in the
laboratory frame) as calculated from the instanton-induced partons within
the ``band''
\begin{eqnarray}
{\rm Sphericity}^{(I)}_{\mid {\rm lab}}&\approx& 0.45,\nonumber\\
{\rm Aplanarity}^{(I)}_{\mid {\rm lab}}&\approx&0.22,\\
{\rm Pseudo\ sphericity}^{(I)}_{\mid {\rm lab}}&\approx&0.90.\nonumber
\end{eqnarray}
Of course, due to the isotropy of the  instanton-induced subprocess,
the sphericity, aplanarity and pseudo-sphericity variables essentially
adopt their maximal values of 1, 1/2 and 1, respectively,
in the $I$-c.m. frame.
Apparently, the pseudo sphericity is least affected by the
Lorentz transformation into the laboratory frame and continues to reflect
the underlying isotropy. Clearly, for a quantitative discussion,
hadronization effects  have to be included\cite{grs}, which may well
wash out somewhat the traces of the underlying isotropy in the
various event shape parameters.

\vspace{0.2cm}
\section{Summary and Outlook}
\noindent
The search for QCD-instanton induced events at HERA is
well worth the effort:
\bigskip

First of all, these ``anomalous'' processes are
predicted to occur {\it within standard QCD}.
Secondly, there is a close analogy to electroweak $B+L$ violating
processes, as was discussed in detail in Sect.~2.
While a promising search for anomalous
electroweak events is only possible in the far future,
the search for manifestations of
QCD-instantons  can start right now,
in deep inelastic $e^\pm p$ scattering at HERA.

Besides summarizing the essence and limitations of the theoretical
calculations involved \cite{bb}, we have presented in Sect.~3 a state
of the art evaluation of the instanton-induced contribution
to the nucleon structure function $F_2$.
It rises strongly with decreasing $x_{\rm Bj}$ and tends
to reach the size of the experimental data around
$x_{\rm Bj} \approx 0.1 \div 0.25$. Unfortunately, due to
inherent uncertainties, the calculation cannot be trusted
anymore for $x_{\rm Bj}\lwig 0.35$, say.
Nevertheless, the trend is very suggestive! However,
enhancements of the inclusive leptoproduction rate
at small $x_{\rm Bj}$ are also expected from other competing
mechanisms like ``(perturbative) Reggeization''.
Therefore, the structure functions appear only of limited
value in searches for ``footprints'' of QCD-instantons.

In Section 4 we have reported first phenomenological results
of our ongoing broad and systematic investigation
of the QCD-instanton induced {\it hadronic final state} \cite{rs}.
Since a calculation of the ``splitting function'' --
associated with the propagation of the current quark in the
instanton background -- is still in progress,
a discussion of expected event rates has to be deferred to
a later stage. In the present analysis we only made use of information
abstracted from our calculation of the instanton-subprocess
observables \cite{rs} along with HERA kinematics.
This was sufficient, however, to obtain important insight into
the expected event topology in the most interesting regime
of small $x_{\rm Bj}$.
\begin{itemize}
\item The ``0th level'' signature to watch out for is a densely
      populated hadronic ``band''  in the
      $(\eta^{\rm lab},\phi^{\rm lab})$-plane
      (c.\,f. Fig.~\ref{f7}), centered at some
      fluctuating
      value $\eta_I^{\rm lab}$. This striking multi-hadron final
      state originates from $8\div 10$ ``semi-hard'' jets
      (c.\,f. Fig.~\ref{f6}),
      always includes strangeness and is characterized by a
      width $\triangle \eta_I^{\rm lab}=\pm 0.9$.
      It directly reflects an underlying instanton-subprocess,
      associated with the formation of an $S$-wave ``fireball'', which
      then decays isotropically into gluons and at least $2n_f-1$ quarks.
      Observables which are particularly sensitive to this event
      structure are e.g. the (transverse) energy flow and the
      so-called pseudo sphericity. The energy flow signature may
      well be less affected by hadronization
      than patterns associated with individual tracks.
\item As a next level of sophistication, we have studied
      kinematical cuts on suitable final-state variables,
      which help to further enhance
      the event topolology and to unfold and/or
      restrict the (Bjorken) variables of the instanton subprocess
     within a theoretically controllable regime, despite small $x_{\rm Bj}$.
     Along these lines one may hope to bridge the
     substantial gap between the kinematical region, where the
     instanton-subprocess cross sections may be theoretically estimated
      \cite{rs}, and
      small $x_{\rm Bj}$ values, $x_{\rm Bj} \lwig O(10^{-3})$,
      where the bulk of HERA data is accumulating at present.
\end{itemize}
In summary, experimental searches for instanton
``footprints'' in the multi-particle final state appear to
be much more promising than searches via the structure functions.

Finally, let us briefly mention some related theoretical and
phenomenological issues presently under study.
Theoretical work is in progress to improve the pre-exponential factors,
affecting quite strongly the predictions for structure functions
and the various subprocess cross sections. This refers in particular to
a more reliable evaluation of the functional determinants \cite{frs} in the
instanton/anti-instanton valley background beyond the
dilute instanton-gas approximation.

Of great importance for further studies of QCD-in\-stan\-ton
phe\-no\-me\-nology is the task of establishing
a convolution form \cite{mrs} of the $\gamma^\ast$-parton multi-particle
cross sections in terms of ``splitting functions'' and
instanton-subprocess cross sections (c.\,f. Eq.~(\ref{convolution})).
Once the ``splitting functions'' have been isolated and calculated,
we hope to come forward with predictions for the rate of
instanton-induced multi-particle events.
We are then ready to study the instanton-induced
multi-particle final state by means of a Monte Carlo based event
generator \cite{grs}. Only after including effects of hadronization
and background will it be possible to address the crucial question:
How many ``anomalous'' events are needed to establish the
``discovery'' of an instanton at HERA?

\vspace{0.2cm}
\section*{Acknowledgements}
We would like to thank
W. Bartel, T. Haas,
M. Kuhlen and A. de Roeck for many useful
suggestions on experimental issues.
Furthermore, we would like to acknowledge helpful discussions with
V. Braun, S. Moch, G. Schuler and C. Wetterich.

\end{document}